\documentclass[10pt]{article}
\usepackage{latexsym}
\usepackage{epsfig}
\usepackage{amsmath}
\usepackage{graphics}
\usepackage{amssymb}
\usepackage{amsthm}
\usepackage{amsopn}
\usepackage{amscd}
\usepackage{fullpage}

\newtheorem{thm}{Theorem}

\numberwithin{equation}{section}
\numberwithin{figure}{section}
\numberwithin{thm}{section}

\begin{document}



\title{Using Simulated Annealing to Calculate the Trembles of Trembling Hand
Perfection}
\begin{center}

\textbf{Stuart McDonald}  \\
School of Economics \\
The University of Queensland \\
Queensland 4072, \\
Australia \\
s.mcdonald@mailbox.uq.edu.au\\
\end{center}

\begin{center}
\textbf{Liam Wagner} \\
Department of Mathematics and \\
St John's College, within \\
The University of Queensland \\
Queensland 4072, Australia \\
LDW@maths.uq.edu.au
\end{center}

\begin{abstract}
\noindent Within the literature on non-cooperative game theory, there have been a
number of algorithms which will compute Nash equilibria.
This paper shows that the family of algorithms known as Markov chain Monte Carlo (MCMC) can be used to calculate Nash equilibria. MCMC is a type of Monte Carlo simulation that relies on Markov chains to ensure its regularity conditions. MCMC has been widely used
throughout the statistics and optimization literature, where variants of
this algorithm are known as simulated annealing. This paper shows that there
is interesting connection between the trembles that underlie the functioning
of this algorithm and the type of Nash refinement known as trembling hand
perfection. This paper shows that it is possible to use simulated annealing to compute this refinement.
\end{abstract}


\section{Introduction}
This paper develops an algorithm to compute a desired type of Nash Equilibrium.
Furthermore we use this algorithm to show existance and uniqness of sensible Nash Equilibrium. Our novel approach to this
problem has been motivated by the number of existance algorithms. The basis of the general
approach of the literature has been to rely on the geometric properties of the equilibrium.

This paper is interested in computing Nash
equilibria that satisfy the type of Nash  of refinement refered to as "trembling hand"
perfection \cite{Selt75} \cite{Selt78}. This paper shows that simulated annealing can be used to compute the above refinement. Simulated annealing is a type of Monte Carlo sampling procedure that relies on Markov chains to ensure its regularity conditions.
Most applications have mainly concentrated on problems of
combinatorial optimization such as routing and packing problems, or problems
from statistical pattern recognition like image processing.

Another well known group of algorithms for calculating Perfect Nash Equilibria are the
trace algorithms of Harsanyi and Selten
\cite{H+S}, where an outcome for the game is selected by ``tracing'' a
feasible path through a family of auxiliary games. The solution progress
along the feasible path is intended to represent the way in which players
adjust their expectations and predictions about the play of the game.

A major limitation of the tracing procedure is that the logarithmic version of
this method, does not always provide a path that traces to a perfect
equilibrium. Harsanyi \cite[p.69]{harsanyi}, has argued that this problem can be
resolved by eliminating all dominated pure strategies before applying the tracing
procedure. However van Damme \cite[p.77]{vDam91} constructs examples which do
not rquire dominated pure strategies in which the tracing procedure yields a
non-perfect equilibrium. Furthermore it was suggested by van Damme that the
inconsistancy lies in the logarithmic control costs. Games which have a control
cost parameter are of normal form so that players may also choose strategies,
incur depending on how well they choose to control their actions.

Another limitation of the tracing procedure it relies on the algeobro-geometric
properties of the equilibrium. This approach has been commonly used throughout
the literature for computing the equilibrium of non-cooperative games. For
example the focus of Lemke and Howson \cite{L+H} for bimatrix
games and the Wilson \cite{Wils71} and Scarf \cite{Scar73} algorithm for the
$N$-person games has also been to utilise the fundamental geometry of games to
calculate equilibrium. In general these approaches to Equilibrium calculation
are computationally expensive. 

However, within game theory there is a history of Monte Carlo methods being
applied to solve non-cooperative games, e.g. starting with Ulam \cite{Ulam50}
in 1954. From the view point of applying global optimization techniques to
infinite games, Monte Carlo simulation has been used by Georgobiani and
Torondzadze as a means of providing Nash equilibria for rectangular games
\cite{GT80}. This is the approach that we will be developing in this paper.

This paper is organised as follows.
The second section of this paper introduces the MCMC algorithm and provides
some discussion of its convergence properties in terms of Markov chain
theory. As a starting point for this discussion the connection between MCMC
sampling techniques and Monte Carlo sampling techniques is explored. The
MCMC algorithms include the Gibbs sampler and the Metropolis algorithm and
are often called simulated annealing. The third section of this paper will
provide a characterization of these algorithms in terms of the trembling
hand of trembling hand perfection. With this in mind, we provide an example
of the use of simulated annealing applied to calculating Nash equilibrium.
In this example the solution leads to equilibria that result from trembling
hand perfection.

\section{A Review of Simulated Annealing}

Monte Carlo simulation has been used extensively for solving complicated
problems that defy an analytic formulation. The main idea behind Monte Carlo
simulation is to either construct a stochastic model that is in agreement
with the actual problem analytically, or to simulate the problem directly.
One problem with Monte Carlo methods is that if the underlying probability
distribution is non-standard, then the convergence of sampled stochastic
process cannot be assured by the SLLN. One way around this is to realize
that a stochastic process can be generated from any process that draws its
samples from the support of underlying distribution. Markov chain Monte
Carlo (MCMC) does this by constructing a Markov chain that uses the
underlying distribution as its stationary distribution. This enables the
simulation of the stochastic process for non-standard distributions, while
ensuring that the SLLN will hold.

As an illustration of the MCMC we will discuss the \emph{Metropolis algorithm} \cite
{MRRTT53}. In this algorithm, each iteration will comprise $h$ updating
steps. Let $X_{t.i}$ denote the state of $X_{i}$ at the end of the $t$th
iteration. For step $i$ of iteration $t+1$, $X_{i}$ is updated using the
Metropolis algorithm. The candidate $Y_{i}$ is generated from a \emph{%
proposal distribution} $q_{i}\left( Y_{i}|X_{t,i},X_{t,-i}\right) $, where $%
X_{t,-i}$ denotes the value of
\begin{equation*}
X_{-i}=\left\{ X_{1},...,X_{i-1},X_{i+1},...,X_{h}\right\}
\end{equation*}
after completing step $i-1$ of iteration $t+1$, i.e.
\begin{equation*}
X_{t,-i}=\left\{ X_{t+1,1},...,X_{t+1,i-1},X_{t.i+1},...,X_{t.h}\right\} ,
\end{equation*}
where the components $X_{t,i+1},...,X_{t,h}$ have yet to be updated and
components $X_{t+1,1},...,X_{t+1,i-1}$ have already been updated. Thus the
proposal distribution of the $i$th component $q_{i}\left( \cdot |\cdot
,\cdot \right) $, generates a candidate for only the $i$th component of $X$.
The candidate is accepted with probability
\begin{equation*}
\alpha \left( X_{-i},X_{i},Y_{i}\right) =\min \left( 1,\frac{\pi \left(
Y_{i}|X_{-i}\right) q\left( X_{i}|Y_{i},X_{-i}\right) }{\pi \left(
X_{i}|X_{-i}\right) q\left( Y_{i}|X_{i},X_{-i}\right) }\right) ,
\end{equation*}
where
\begin{equation*}
\pi \left( X_{i}|X_{-i}\right) =\frac{\pi \left( X\right) }{\int \pi \left(
X\right) dX_{.i}}
\end{equation*}
is the full conditional distribution for $X_{i}$ under $\pi \left( \cdot
\right) $. If $Y_{.i}$ is accepted, then $X_{t+1,i}=Y_{i}$; otherwise $%
X_{t+1,i}=X_{t,i}$. For this reason $\alpha \left(
X_{.-i},X_{.i},Y_{.i}\right) $ is known as the \emph{Metropolis criterion}.

One of the disadvantages of this algorithm is the complexity of the
Metropolis criterion\emph{\ }$\alpha \left( X_{.-i},X_{.i},Y_{.i}\right) $.
In practice $\alpha \left( X_{.-i},X_{.i},Y_{.i}\right) $ often simplifies
considerably, particularly when $\pi \left( \cdot \right) \,$derives from a
conditional independence model \cite{Gilks96} \cite{Rob96}. However, the
single component Metropolis algorithm has the advantage of employing the
full conditional distributions for $\pi \left( \cdot \right) $ and Besag
\cite{Besag74} has shown that $\pi \left( \cdot \right) $ will be uniquely
determined by its full conditional distribution. As a result $\alpha \left(
X_{.-i},X_{.i},Y_{.i}\right) $ will generate samples from a unique target
distribution $\pi \left( \cdot \right) $.

An alternative approach for constructing a Markov chain with a stationary
distribution $\pi \left( \cdot \right) ,$ that provides a generalization of
the approach suggested by Metropolis et al. \cite{MRRTT53}, has been
suggested by Hastings \cite{Hast70}. At each point in time $t$, the next
state $X_{t+1}$ is chosen by first sampling a candidate point $Y$ from a
proposal distribution $q\left( \cdot |X_{t}\right) $. The candidate point $Y$
is then accepted in accordance with the criterion
\begin{equation*}
\alpha \left( X,Y\right) =\min \left( 1,\frac{\pi \left( Y\right) }{\pi
\left( X\right) }\right) .
\end{equation*}
Under this criterion, if the candidate point is accepted, then $X_{t+1}=Y$,
otherwise $X_{t+1}=X_{t}$. The main difference between this algorithm and
the one proposed by Metropolis et al. \cite{MRRTT53}, is that the \emph{%
Metropolis-Hastings algorithm}, as it is named, assumes that the proposal
distributions are symmetric, i.e. $q\left( Y|X\right) =q\left( X|Y\right) $.
The Metropolis-Hastings algorithm is therefore ruled out for higher
dimensional problems, as these problems generally have little symmetry. The
main advantage of the Metropolis-Hastings algorithm is that proposal
distribution has no impact on the decision criterion, and therefore will not
impact on the convergence of this algorithm towards the stationary
distribution $\pi \left( \cdot \right) $.

To provide a fuller explanation, the transition kernel of the
Metropolis-Hastings algorithm is given by
\begin{equation}
\begin{split}
&P\left( X_{t+1}|X_{t}\right) =q\left( X_{t+1}|X_{t}\right) \alpha \left(
X_{t},X_{t+1}\right) \\
&+I\left( X_{t+1}=X_{t}\right) \left[ 1-\int q\left( Y|X_{t}\right) \alpha
\left( X_{t},Y\right) dY\right] ,
\end{split}
\end{equation}

where $I\left( \cdot \right) $ is the indicator function. From $\alpha
\left( X_{t},X_{t+1}\right) $, we can see that
\begin{equation*}
\begin{split}
&\pi \left( X_{t}\right) q\left( X_{t+1}|X_{t}\right) \alpha \left(
X_{t},X_{t+1}\right) =\\
&\pi \left( X_{t+1}\right) q\left( X_{t}|X_{t+1}\right)
\alpha \left( X_{t+1},X_{t}\right) .
\end{split}
\end{equation*}
This implies that
\begin{equation*}
\pi \left( X_{t}\right) P\left( X_{t+1}|X_{t}\right) =\pi \left(
X_{t+1}\right) P\left( X_{t}|X_{t+1}\right) .
\end{equation*}
Integrating both sides of this equation, we get
\begin{equation*}
\int \pi \left( X_{t}\right) P\left( X_{t+1}|X_{t}\right) dX_{t}=\pi \left(
X_{t+1}\right) .
\end{equation*}
This equation states that if $X_{t}$ is drawn from $\pi $, then so must $%
X_{t+1}$. In other words, once one sample value has been obtained from the
stationary distribution, then all subsequent samples must be drawn from the
same distribution.

This is only a partial justification of the Metropolis-Hastings algorithm. A
full proof requires that $P^{\left( t\right) }\left( X_{t}|X_{0}\right) $
converges on the stationary distribution. For a heuristic justification of
this result, it can be noted that this distribution will depend only on the
starting value $X_{0}$, therefore the proof must show that Markov chain
gradually forgets its starting point, and converges on a unique stationary
distribution. Thus, after a sufficiently long \emph{burn-in} of $m$
iterations, points $\left\{ X_{t};t=m+1,\,...,n\right\} $ will be dependent
sample approximations of the stationary distribution. Hence the \emph{%
burn-in sample} is usually discarded when calculating the ergodic mean for $%
f\left( X\right) $%
\begin{equation*}
\bar{f}=\frac{1}{m-n}\sum_{t=m}^{n}f\left( X_{t}\right) .
\end{equation*}

\section{Trembling Hand Algorithm}

\subsection{A MCMC Algorithm for Computing Perfect Equilibria in
Strategic Games}

In this sub-section we provide an algorithm for computing a perfect
equilibrium for a strategic game and show that this algorithm
provides a sequence of perturbed mixed strategies that will
eventually converge on perfection. The basic idea is to construct
select a Markov chain and then use this Markov to deliver a Nash
equilibrium via Markov chain approximation. The trick is to
nominate the appropriate Markov chain with the most suitable
convergence properties to deliver convergence of the sequence
completely mixed Nash equilibria of perturbed games or $\varepsilon
$-perfect equilibria to a perfect equilibrium. This is the
objective that is undertaken in this section.

Consider an $n$-person game in strategic form $G=\left( N,\left(
S_{i}\right) _{i\in N},\left( u_{i}\right) _{i\in N}\right) $ in which $%
N=\left\{ 1,...,n\right\} $ is the player set, each player $i\in N$ has a
finite set of pure strategies $S_{i}=\left\{ s_{i1},...,s_{ik_{i}}\right\} $
and a pay-off function $u_{i}:\times _{i\in N}S_{i}\rightarrow \mathbb{R}$
mapping the set of pure strategy profiles $\times _{i\in N}S_{i}$ into the
real number line.

In the strategic game $G$, for each player $i\in N$ there is a set of
probability measures $\Delta _{i}$ that can be defined over the pure
strategy set $S_{i},$ this is player $i$'s mixed strategy set. The elements
of the set $\Delta _{i}$ are of the form $p_{i}:S_{i}\rightarrow \left[
0,1\right] $ where $\sum_{j=1}^{k_{i}}p_{ij}=1,$ with $p_{ij}=p\left(
s_{ij}\right) ,$ i.e. $\Delta _{i}$ is isomorphic to the unit simplex.

We denote the elements of the space of mixed strategy profiles $\times
_{i\in N}\Delta _{i}$ by $p=\left( p_{1},...,p_{n}\right) ,$ where $%
p_{i}=\left( p_{i1},...,p_{ik_{i}}\right) \in \Delta _{i}$. As is the
convention we use the following short-hand notation $p=\left(
p_{i},p_{-i}\right) $, where $p_{-i}$ denotes the other components of $p$.

For each player $i$, the pay-off function $u_{i}:\times _{i\in N}\Delta
_{i}\rightarrow \mathbb{R}$ can be extended to the domain of mixed strategy
profiles $\times _{i\in N}\Delta _{i}$. The pay-off function for each player
$i\in N$ will be defined as follows $u_{i}\left( p_{i},p_{-i}\right)
=\sum_{j=1}^{k_{i}}p_{ij}u_{i}\left( s_{ij},p_{-i}\right) $. A mixed
strategy $p\in $ $\times _{i\in N}\Delta _{i}$ is \textbf{Nash equilibrium}
of the strategic game $G$, if for all players $i\in N$ and all $%
p_{i}^{\prime }\in \Delta _{i}$
\begin{equation}
u_{i}\left( p_{i},p_{-i}\right) \geq u_{i}\left( p_{i}^{\prime
},p_{-i}\right) .
\end{equation}

Suppose that as well there being a positive probability $p_{ij}$ of a player
$i$ selecting a pure strategy s$_{ij}\in S_{i}$, there is a small
probability $\varepsilon _{ij}$ that the pure strategy $s_{ij}$ will be
chosen by $i$ out of error. In the case where player $i$ selects his $j$th
pure strategy $s_{ij}$ by mistake, the probability of doing so is given by $%
q_{ij}$. The total probability of player $i$ selecting a pure strategy s$%
_{ij}\in S_{i}$ is then given by
\begin{equation}
\hat{p}_{ij}=\left( 1-\varepsilon _{ij}\right) p_{ij}+\varepsilon
_{ij}q_{ij}.
\end{equation}

It can be seen that in this case, the total probability of player $i$
selecting a pure strategy s$_{ij}\in S_{i}$ will be bounded below by
\begin{equation}
\hat{p}_{ij}\geq \varepsilon _{ij}q_{ij}.
\end{equation}
Equating $\eta _{ij}=\varepsilon _{ij}q_{ij}$ we can see that this condition
can be rewritten as
\begin{equation}
\hat{p}_{ij}\geq \eta _{ij}\quad \forall \,s_{ij}\in S_{i}\text{ and }i\in N,
\end{equation}
with
\begin{equation}
\sum_{j=1}^{k_{i}}\eta _{ij}<1\quad \forall \,i\in N.
\end{equation}

This leads to the definition of a perturbed game $\left( G,\eta \right) $ as
a finite strategic game derived from the strategic game $G$, in which each
player $i$'s mixed strategy set is the set of completely mixed strategies
for player $i$ constrained by the probability of making an error
\begin{equation}
\Delta _{i}\left( \eta _{i}\right) = p_{i}=\left\{ \left(
p_{i1},....,p_{ik_{i}}\right) \in \Delta _{i};p_{ij}\geq \eta _{ij}\, \text{and }
\sum\nolimits_{j=1}^{k_{i}}\eta _{ij}<1\right\}  
\end{equation}
A mixed strategy combination $p\in \times _{i\in N}\Delta _{i}\left( \eta
_{i}\right) $ is a Nash equilibrium of the perturbed game $\left( G,\eta
\right) $ iff the following condition is satisfied
\begin{equation}
u_{i}\left( s_{ij},p_{-i}\right) <u_{i}\left( s_{il},p_{-i}\right) \text{
then }p_{ij}=\eta _{ij},\quad \forall \,s_{ij}\text{,\thinspace }s_{il}\in
S_{j}.
\end{equation}

A mixed strategy $p\in $ $\times _{i\in N}\Delta _{i}$ is a \textbf{perfect
equilibrium} in the strategic game $G$ if there exists a sequence of
completely mixed strategy profiles $\left\{ p^{k}\right\} _{k=1}^{\infty }$
where $\lim_{k\rightarrow \infty }p^{k}=p$, and for every player $i\in N$
and for every $p_{i}^{\prime }\in \Delta _{i}$%
\begin{equation}
u_{i}\left( p_{i},p_{-i}^{k}\right) \geq u_{i}\left( p_{i}^{\prime
},p_{-i}^{k}\right) \quad \forall \,k=1,2,....
\end{equation}
In terms of our definition of a perturbed game, a mixed strategy is a
perfect equilibrium iff there exist some sequences $\left\{ \eta ^{k}=\left(
\eta _{1}^{k},...\eta _{n}^{k}\right) \right\} _{k=1}^{\infty }$ and $%
\left\{ p^{k}=\left( p_{1}^{k},...p_{n}^{k}\right) \right\} _{k=1}^{\infty }$
such that

\begin{enumerate}
\item  each $\eta ^{k}>0$ and $\lim_{k\rightarrow \infty }\eta _{k}=0$,

\item  each $p^{k}$ is a Nash equilibrium of a perturbed game equilibrium $%
\left( G,\eta ^{k}\right) $, and

\item  $\lim_{k\rightarrow \infty }p^{k}=p$ where for every player $i\in N$
and for every $p_{i}^{\prime }\in \Delta _{i}$%
\begin{equation}
u_{i}\left( p_{i},p_{-i}^{k}\right) \geq u_{i}\left( p_{i}^{\prime
},p_{-i}^{k}\right) \quad \forall \,k=1,2,....
\end{equation}
\end{enumerate}

An alternative definition of perfection has been made Myerson
\cite[pp 75--76]{Myers78} and is based on the idea that every pure strategy
in a player's set of pure strategies has associated with it a small positive
probability of at least $\varepsilon >0,$ but on strategies that are best
responses have associated probabilities greater that $\varepsilon .$ More
formally, for any player $i\in N$ a mixed strategy $p_{i}\in \Delta _{i}$ is
an $\varepsilon $\textbf{-perfect equilibrium} iff it is completely mixed
and
\begin{equation}
u_{i}\left( s_{ij},p_{-i}\right) <u_{i}\left( s_{il},p_{-i}\right) \text{
then }p_{ij}\leq \varepsilon ,\text{\quad }\forall \,s_{ij}\text{,\thinspace
}s_{il}\in S_{j}.
\end{equation}
Unlike Nash equilibria of perturbed games, the $\varepsilon $-perfect
equilibria of a game $G$ will not necessarily be one of its Nash equilibria.
However, Myerson does show that $p=\left( p_{1},...,p_{n}\right) \in \times
_{i\in N}\Delta _{i}$ will be a perfect equilibrium iff

\begin{enumerate}
\item  each $\varepsilon ^{k}>0$ and $\lim_{k\rightarrow \infty }\varepsilon
^{k}=0$,

\item  each $p^{k}$ is an $\varepsilon ^{k}$-perfect equilibrium of the game
$G$, and

\item  $\lim_{k\rightarrow \infty }p_{i}^{k}=p_{i}$ for every player $i\in
N. $
\end{enumerate}

The starting basis for the MCMC algorithm for calculating
perfection will be to follow Myerson by constructing a sequence of
$\varepsilon $-perfect equilibria for the strategic game $G$. As
stated above, we know that for the
strategic game $G$, $p\in \times _{i\in N}\Delta _{i}$ is an $\varepsilon $%
-perfect equilibrium iff for each player $i\in N$, $p_{i}\in \Delta
_{i}$ is a completely mixed strategy and
\begin{equation}
\begin{split}
&u_{i}\left( s_{ij},p_{-i}\right) <u_{i}\left( s_{il},p_{-i}\right)
\text{ then }p_{ij}\leq \varepsilon,\\
&\text{\quad }\forall
\,s_{ij}\text{,\thinspace }s_{il}\in S_{j}.
\end{split}
\end{equation}

Following Myerson \cite[p 79]{Myers78} we define the following set
of mixed strategies for each player $i\in N$
\begin{equation}
\Delta _{i}^{*}=\left\{ p_{i}\in \Delta _{i};p_{ij}\geq \delta
\;\,\forall \,s_{ij}\in S_{i}\right\} ,
\end{equation}
where
\begin{equation}
\delta =\frac{1}{m}\varepsilon ^{m},\quad 0<\varepsilon <1
\end{equation}
with $m=\max_{i\in N}\left| S_{i}\right| $. We then define a
point-to-set mapping $F_{i}:\times _{i\in N}\Delta
_{i}^{*}\rightarrow \Delta _{i}^{*}$ to be a family of completely
mixed distributions contained in $\Delta _{i}^{*}$
\begin{equation}
\begin{split}
&F_{i}\left( p_{1},...,p_{n}\right) =\left\{ p_{i}^{*}\in \Delta
_{i}^{*};u_{i}\left( s_{ij},p_{-i}\right) <u_{i}\left(
s_{il},p_{-i}\right)\right.\\
&\left.\text{ then }p_{ij}\leq \varepsilon ,\text{\quad }\forall \,s_{ij}\text{%
,\thinspace }s_{il}\in S_{j}\right\}
\end{split}
\end{equation}

If we then define, for each player $i\in N$, a mixed strategy
\begin{equation}
p_{il}^{*}=\frac{e^{\rho \left( s_{ij}\right)
}}{\sum_{l=1}^{k_{i}}e^{\rho \left( s_{il}\right) }},
\end{equation}
where
\begin{equation}
\rho \left( s_{ij}\right) =\left| \left\{ s_{il}\in
S_{i};u_{i}\left( s_{ij},p_{-i}\right) <u_{i}\left(
s_{il},p_{-i}\right) \text{ and }p\in \times _{i\in N}\Delta
_{i}^{*}\right\} \right|
\end{equation}
Then it can be seen that $p_{i}^{*}\in F_{i}\left(
p_{1},...,p_{n}\right) $ will be non-empty. As each $F_{i}\left(
p_{1},...,p_{n}\right) $ will a finite collection of linear
inequalities, they will also be closed convex sets. In addition
each $F_{i}\left( p_{1},...,p_{n}\right) $, by the continuity of
the pay-off function $u_{i}\left( s_{ij},\cdot \right) ,$ will also
be upper semi-continuous.

As a consequence the mapping $F:\times _{i\in N}\Delta
_{i}^{*}\rightarrow \times _{i\in N}\Delta _{i}^{*}$ satisfies all
the conditions of the Kakutani Fixed Point Theorem. In other words
there exists some completely mixed strategy $p_{\varepsilon }\in
\times _{i\in N}\Delta _{i}^{*}$ such
that $p_{\varepsilon }$ is an $\varepsilon $-perfect equilibrium of $G$. As $%
\times _{i\in N}\Delta _{i}$ is compact, the sequence $\varepsilon
$-perfect
equilibria $p_{\varepsilon }\rightarrow $ $p$ as $\varepsilon \rightarrow 0$%
, where $p$ is the perfect equilibrium of $G$.

An alternative route to the same result can be arrived at as
follows using an argument based on the convergence properties
Markov chain.

\begin{thm}
For any normal form game $G=\left( N,\left( S_{i}\right) _{i\in
N},\left( u_{i}\right) _{i\in N}\right) $, it is possible to define
a MCMC algorithm such that its transition probabilities will
converge to a perfect equilibrium as long as the following
conditions hold:

\begin{enumerate}
\item  if $u_{i}\left( s_{ij},p_{-i}^{k}\right) -u_{i}\left(
s_{il},p_{-i}^{k}\right) \geq 0$ then accept, where $p_{-i}^{k}$ is
the tuple mixed strategies selected on the $k$th iteration;

\item  otherwise, accept if probability $\exp \left( \frac{u_{i}\left(
s_{il},p_{-i}^{k}\right) -u_{i}\left( s_{il},p_{-i}^{k}\right)
}{T}\right)
>\varepsilon ,$ where $\varepsilon \sim U\left[ 0,1\right] ;$ and

\item  in addition it can be seen that for all $s_{ij}$ and $s_{il}\in S_{i}$
such that $u_{i}\left( s_{ij},p_{-i}^{k}\right) <u_{i}\left(
s_{il},p_{-i}^{k}\right) $, $\alpha _{jl}^{i}\left( T\right)
\rightarrow 0$ as $T\rightarrow \infty $.
\end{enumerate}
\end{thm}

\noindent
\proof%
For each player $i\in N$, there will be a collection these subsets
\begin{equation}
N_{ij}=\left\{ s_{il}\in S_{i};u_{i}\left( s_{ij},p_{-i}\right)
<u_{i}\left( s_{il},p_{-i}\right) \text{ and }p\in \times _{i\in N}\Delta _{i}^{*}\right\}
\end{equation}
of $i$'s pure strategy space $S_{i}$. The collection of these sets
will referred to as player $i$'s local neighborhood structure. What
we would like to do is for any two pure strategies
$s_{ij}$,$\,s_{il}\in S_{i}$ define a path from $s_{ij}$ to
$s_{il}$ such that
\begin{equation}
s_{ij_{1}}\in N_{ij},s_{ij_{2}}\in N_{ij_{1}},...,s_{il}\in
N_{ij_{m}}.
\end{equation}

In order to do this, we observe that the point-set mapping defined
by the set
\begin{equation}
F_{i}\left( p_{1},...,p_{n}\right) =\left\{ p_{i}^{*}\in \Delta
_{i}^{*};u_{i}\left( s_{ij},p_{-i}\right) <u_{i}\left(
s_{il},p_{-i}\right)\text{ then }p_{ij}\leq \varepsilon ,\text{\quad }\forall \,s_{ij}\text{%
,\thinspace }s_{il}\in S_{i}\right\}
\end{equation}
is a collection homogenous transition probabilities $S_{i}$
\begin{equation}
p_{jl}^{i}\left( k\right) =\Pr \left\{ s_{i}\left( k\right)
=s_{il}|s_{i}\left( k-1\right) =s_{ij}\right\} =\Pr \left\{
s_{il}|s_{ij}\right\} .
\end{equation}
Further more we can see that these transition probabilities have
the Markov property, i.e. given the path from $s_{ij}$ to $s_{il}$
such that
\begin{equation}
s_{ij_{1}}\in N_{ij},s_{ij_{2}}\in N_{ij_{1}},...,s_{il}\in
N_{ij_{m}}.
\end{equation}
the conditional probability
\begin{equation}
\begin{split}
&\Pr \left\{s_{il}s_{ij_{1}},s_{ij_{2}},...s_{ij_{m}},s_{ij}\right\} \\
&=\Pr
\left\{ s_{il}|s_{ij_{m}}\right\} \Pr \left\{
s_{ij_{m}}|s_{ij_{m-1}}\right\} ..\Pr \left\{
s_{ij_{2}}|s_{ij_{1}}\right\}
\end{split}
\end{equation}

We define the following generating probability for the Markov chain
for each
player $i\in N$%
\begin{equation}
g_{jl}^{i}=\left\{
\begin{array}{l}
\frac{1}{\rho \left( s_{ij}\right) }\text{,\quad if }s_{il}\in
N_{ij} \\ 0,\quad \quad \;\;\text{otherwise},
\end{array}
\right.
\end{equation}
where
\begin{equation}
\rho \left( s_{ij}\right) =\left| \left\{ s_{il}\in
S_{i};u_{i}\left( s_{ij},p_{-i}\right) <u_{i}\left(
s_{il},p_{-i}\right)
\text{ and }p\in \times _{i\in N}\Delta
_{i}^{*}\right\} \right| .
\end{equation}
We now introduce the following acceptance probability
\begin{equation}
\begin{split}
\alpha _{jl}^{i}\left( T\right) &=\left\{ 1,\exp \left(
\frac{u_{i}\left(
s_{ij},p_{-i}^{k-1}\right) -u_{i}\left( s_{il},p_{-i}^{k-1}\right) }{T}%
\right) \right\} ,\\
&T>0
\end{split}
\end{equation}
where $T$ is a control parameter. This last condition implies that

\begin{enumerate}
\item  if $u_{i}\left( s_{ij},p_{-i}^{k}\right) -u_{i}\left(
s_{il},p_{-i}^{k}\right) \geq 0$ then accept, where $p_{-i}^{k}$ is
the tuple mixed strategies selected on the $k$th iteration;

\item  otherwise, accept if probability $\exp \left( \frac{u_{i}\left(
s_{il},p_{-i}^{k}\right) -u_{i}\left( s_{il},p_{-i}^{k}\right)
}{T}\right)
>\varepsilon ,$ where $\varepsilon \sim U\left[ 0,1\right] ;$ and

\item  in addition it can be seen that for all $s_{ij}$ and $s_{il}\in S_{i}$
such that $u_{i}\left( s_{ij},p_{-i}^{k}\right) <u_{i}\left(
s_{il},p_{-i}^{k}\right) $, $\alpha _{jl}^{i}\left( T\right)
\rightarrow 0$ as $T\rightarrow \infty $.
\end{enumerate}

Given theses three conditions we can now see that the following
will hold:

\begin{itemize}
\item  We know that under this acceptance criterion as $k\rightarrow \infty $
The transition probability matrix $p_{i}^{k}$ of the homogenous
Markov chain generated by the game $G$ will converge on a
stationary distribution $\pi \left( T\right) $ as $k\rightarrow
\infty $.
\begin{equation}
p_{i}^{k}\rightarrow \pi _{i}\left( T\right) =\frac{e^{-C\left( i\right) /T}%
}{\sum_{k\in E}e^{-C\left( k\right) /T}}
\end{equation}
and as $T\rightarrow \infty $
\begin{equation}
\pi _{i}\left( T\right) =\left\{
\begin{array}{l}
\frac{1}{\left| N_{i}\right| }\quad \text{ if }i\in H \\ 0\quad
\quad \;\text{otherwise}
\end{array}
\right.
\end{equation}
where
\begin{equation}
N_{i}=\left\{ s_{il}\in S_{i};u_{i}\left( s_{ij},p_{-i}\right)
<u_{i}\left( s_{il},p_{-i}\right) ,p_{i}=0\right\} .
\end{equation}
(See van Laarhoven and Aarts \cite[p.22--25]{LA} for the proof of
this last statement.)

\item  The transition probability matrix $p_{i}^{k}$ satisfies Myerson's
definition of an $\varepsilon $-perfect equilibria and as Myerson
has shown, the fixed point that this sequence converges on is also
a perfect
equilibrium.%
\endproof%
\end{itemize}

\section{An Application to Extensive Form Games}

There are problems with viewing the existence of Nash equilibria as an end
in itself. The most immediate problem with this has been the possible large
number of Nash equilibria that can be found for any game, together with the
likelihood that not all of these Nash equilibria will be reasonable in some
sense. One way around this is to view the decision process of each agent
participating in the game from a decision theoretic perspective. From this
viewpoint, only those equilibria that can be found by backwards induction
will be self-enforcing. This leads to a technique for strategy space
reduction by iteratively removing strategies that lead to outcomes that are
not \emph{strongly dominated}. As shown by Kuhn \cite[Corollary 1]{Kuhn53},
under the assumption of perfect information, this leads to a recursion that
is equivalent to the Bellman equation of dynamic programming.

An alternative to this is to construct a recursion that iteratively
eliminates \emph{weakly dominated strategies}. However, the removal of
weakly dominated strategies can lead to the elimination of strategy profiles
that would otherwise provide suitable outcomes if only strongly dominated
strategies were to have been removed. From the viewpoint of this paper these
recursive strategy space reduction techniques can be considered to be an
algorithm that reduces the size of a game, making equilibrium selection
easier. However, these iterative reduction techniques becomes unwieldy once
the assumption of perfect information is relaxed and information sets
contain more than one node of the game tree.

This has led to a number of refinements to the definition of Nash
equilibrium. Among the first of these was the notion of \emph{subgame
perfection} \cite{Selt75}, which removes strategies that are not optimal for
every subgame of a extensive game's game tree. However, Selten \cite{Selt75}
has shown that subgame perfection can also prescribe non-optimizing
behaviour at information sets that are not reached when the equilibrium is
played. This is because the expected payoff for the player whose information
set is not reached will not depend on their own strategy. As a result every
strategy will maximize their payoff. As van Damme \cite[p. 8--9]{vDam91}
states, that this can be removed if the equilibrium prescribes a choice, at
each information set that is a singleton, that maximizes the expected payoff
after the information set. The problem is that not all subgame perfect
equilibria satisfying this criteria are sensible.


Another approach which was suggested by Selten \cite{Selt75}, was to eliminate
``unreasonable'' subgame perfect equilibria by allowing the possibility of
``mistakes'' or ``trembles'' on the part of decision makers. In this way,
isolated information sets are removed, as every information set can now be
reached with positive probability. The other advantage of trembling hand
perfection is that, unlike subgame perfection, it can be applied directly to
the normal form of any game. Although, as van Damme shows, the perfect
equilibria of a game's strategic and extensive forms need not coincide. An
equivalence relationship holds for only the \emph{agent normal form }and
extensive form of any game \cite{Selt75}. This is because the agent normal
form of any game views each node of the game tree, of the extensive form of
the game, as a player in the game. As a consequence each player represents
an information set held by the player and will have an identical payoff
function to the player.

As was shown by Selten \cite{Selt75}, the perfect equilibria of a game's
strategic and extensive forms need not coincide. However he showed that an
equivalence relationship holds between the equilibria of any extensive game
and its associated \emph{agent normal form }\cite{Selt75}. This is because
the agent normal form of any game views each node of the game tree, of the
extensive form of the game, as a player in the game. As a consequence each
player represents an information set held by the player and will have an
identical pay-off function to the player.

We let $\Gamma ^{e}$ define an extensive game consisting of a set of $n$
players, a game tree $K=\left( T,R\right) $ consisting of a set of nodes $T$
and a binary relation $R$ which is a partial ordering on the set of nodes.
The nodes of the game tree are classified as either non-terminal or terminal
according to whether or not their are succeeding nodes in the game tree. The
partial ordering is used to define a path of successive nodes. The
non-terminal nodes of the game tree are partitioned into the sets $%
P_{0},P_{1},...,P_{n}$ that specify the moves associated with each player,
with $P_{0}$ being the partition associated with random moves that are not
associated with any player. All of the non-terminal nodes is the information
partition $U=$ $\left( U_{1},....,U_{n}\right) $, where each set $U_{i}$ is
a partition of $P_{i}$ into information sets, such that all nodes within an
information set $u\in U_{i}$ have the same number of immediate successors
and path intersects an information set at most once. Under the assumption of
perfect information each information set $u\in U_{i}$ will be a singleton.
This paper will assume \emph{imperfect information} -- this implies that if
the information set $u\in U_{i}$ contains a node $x\in P_{i}$, player $i$
will not be able to distinguish other nodes contained in this information
set based on information possessed when moving to $x$. Throughout this paper
it will also be assumed that \emph{complete information} is present -- i.e.
each player has \emph{perfect recall} and will remember everything from
earlier in the game, including their own moves.

Associated with each random move is a probability distribution $p$. The
payoffs associated with the set of terminal points $Z$ of the game tree are
denoted by the $n$-tuple $r=\left( r_{1},...,r_{n}\right) $, where each
player's payoff is a function of the terminal points $r_{i}\left( z\right) $%
, $z\in Z$. With the information partition $U$ a choice set $C=\left\{
C_{u}:u\in \cup _{i=1}^{n}U_{i}\right\} $ can be defined, where each $C_{u}$
is a partition of the union of sets of successors $S\left( x\right) =\left\{
y;x\in P\left( y\right) \right\} $ for each $x\in u$: $\cup _{x\in u}S\left(
x\right) $. The interpretation is that if player $i$ takes the choice $c\in
C_{u}$ at information set $u$ $\in U_{i}$ , then if $i$ is at $x\in u$, the
next node reached is the element of $S\left( x\right) $ contained in $c$.
Under the assumption of imperfect information and perfect recall, a
probability distribution $b_{i}$ is assigned on $C_{u}$ to each information
set $u\in U_{i}.$ This distribution $b_{i}$ is a behavioural strategy, with
the set of all these strategies for player $i$ defined by $B_{i}$. The
profile of all players behavioural strategies is denoted by $b\in B:=\times
_{i=1}^{n}B_{i}$, where $B$ is the set of all behavioural strategy
combinations. The probability of a particular realization of the game $%
\Gamma ^{e}$ is denoted by $\mathbb{P}_{b}\left( z\right) $.

The definition of perfect equilibrium we will use is based Selten \cite
{Selt75} and Friedman \cite{Fried91}. Kuhn \cite{Kuhn53} has shown that
these behavioural and mixed strategies are realization equivalent.
Therefore, for an extensive form game $\Gamma ^{e}$ we let $\Gamma =\left(
S,R\right) $ define its strategic form representation, with $S$ denoting the
set of all mixed strategy profiles. The payoff profile $R$ is an $n$-tuple,
where the $i$th element is defined as
\begin{equation*}
R_{i}=\sum_{z\in Z}\Bbb{P}_{b}\left( z\right) r_{i}\left( z\right) .
\end{equation*}
A perturbed game of $\Gamma $ is defined by $\left( \Gamma ,\eta \right) $,
where $\eta $ is a mapping that assigns to every choice in $\Gamma $ a
positive number $\eta _{c}$ such that
\begin{equation*}
\sum_{c\in C_{u}}\eta _{c}<1
\end{equation*}
for every information set $u$. An equilibrium point $b$ of the strategic
game $\Gamma $ is a perfect equilibrium if $b$ is a limit point of a
sequence $\left\{ b\left( \eta \right) \right\} $ as $\eta \rightarrow 0$,
where each $b\left( \eta \right) $ is an equilibrium points of the
associated perturbed game $\left( \Gamma ,\eta \right) $.

The algorithm is constructed using a simulated annealing algorithm found in
van Laarhoven and Aarts \cite[p. 10]{LA}. The pseudo-code for this algorithm
is given below:

\begin{itemize}
\item[ ]  begin

\item[ ]  \textbf{Intitialize};

\item[ ]  $M:=0$;

\item[ ]  repeat

\begin{itemize}
\item[ ]  repeat

\begin{itemize}
\item[ ]  \textbf{Perturb}(config. $i\rightarrow j$, $\Delta R_{ij}()$) for
player 1;

\item[ ]  if $\left( \Delta R_{ij}\geq 0\right) $ then accept

\begin{itemize}
\item[ ]  elseif $\left( \exp \left( \frac{-\Delta R_{ij}}{c}\right)
>rand\left[ 0,1\right) \right) $ then accept;
\end{itemize}

\item[ ]  if accept then \textbf{Update}(config. $j$);

\item[ ]  \textbf{Perturb}(config. $i\rightarrow j$, $\Delta R_{ij}()$) for
player $n$;

\item[ ]  if $\left( \Delta R_{ij}\geq 0\right) $ then accept

\begin{itemize}
\item[ ]  elseif $\left( \exp \left( \frac{-\Delta R_{ij}}{c}\right)
>rand\left[ 0,1\right) \right) $ then accept;
\end{itemize}

\item[ ]  if accept then \textbf{Update}(config. $j$);
\end{itemize}

\item[ ]  until \textbf{equilibrium is approached sufficiently closely};

\item[ ]  $c_{M+1}:=f\left( c_{M}\right) $;

\item[ ]  $M:=M+1;$
\end{itemize}

\item[ ]  until \textbf{stop criterion = true;}

\item[ ]  end
\end{itemize}

\noindent The energy function differential for this algorithm is defined as
follows:
\begin{equation*}
\Delta R_{ij}=R_{j}-R_{i,}\quad i<j
\end{equation*}
where the $R_{i}$ are the expected pay-off functions for each player
participating in the perturbed game. The temperature function $c$ controls
the trembles and is updated by the decrement rule
\begin{equation*}
c_{M+1}=\alpha \cdot c_{M},\quad 0<\alpha <1,\,M=1,2,...\,\text{.}
\end{equation*}

\begin{figure}[!ht]
\begin{center}\label{tree}
\scalebox{0.5}{\epsfig{file=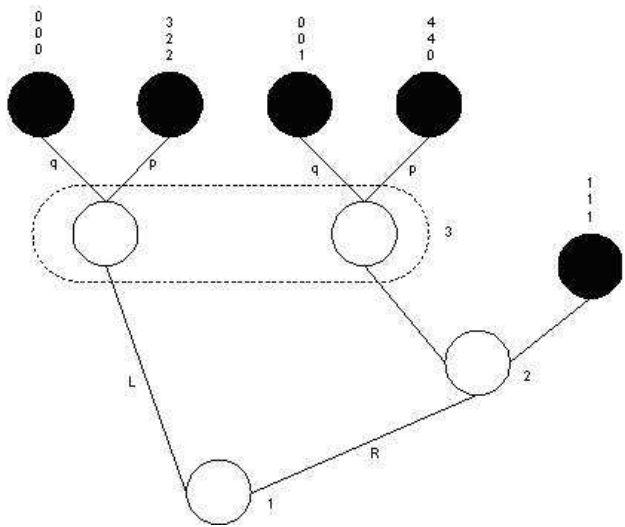
,angle=0,width=\linewidth}}
\caption{Selten's Horse Game Tree}
\end{center}
\end{figure}

We apply it to the following example taken from Friedman \cite[p. 51]
{Fried91}. This example is based on the three player extensive form game
used by Selten \cite{Selt75} to illustrate the existence of perfect
equilibrium. The game tree is defined as follows in Figure \ref{tree}
\cite[p. 50]{Fried91}.

This game possesses both a perfect equilibrium as well as ``non-sensical''
subgame perfect equilibria. The perfect equilibrium for this extensive form
game is defined via the perturbed pay-off functions:

\begin{eqnarray*}
R_{1} &=&\alpha _{1}(1-\varepsilon _{2}-3\varepsilon _{3}+4\varepsilon
_{2}\varepsilon _{3})+3\varepsilon _{3} \\
R_{2} &=&2\varepsilon _{3}(2-\varepsilon _{1})+\alpha _{2}(1-\varepsilon
_{1}-4e_{3}+4\varepsilon _{1}\varepsilon _{3}) \\
R_{3} &=&1-\varepsilon _{1}+\alpha _{3}(2\varepsilon _{1}-\varepsilon
_{2}+\varepsilon _{1}\varepsilon _{2}),
\end{eqnarray*}
where the $\alpha _{i}$ are the mixed strategies and $\varepsilon _{i}$ are
errors defined for $i=1,2,3$. Letting the errors approach zero, it can be
seen that perfect equilibrium is defined by $\left( 1,1,0\right) $.

The results of the simulation are shown below in Figure \ref{payoffs}
and indicate convergence to the trembling hand perfect
equilibrium.
\begin{figure}[!ht]
\begin{center}
\scalebox{0.5}{\epsfig{file=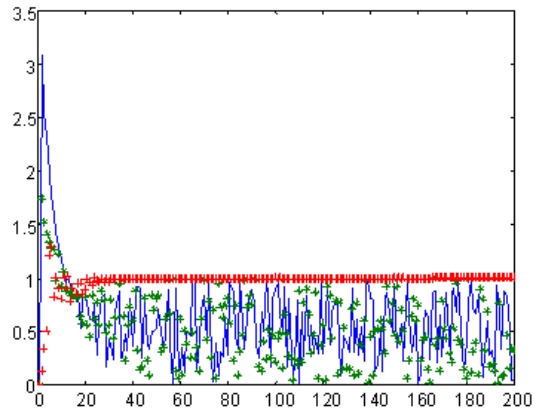
,angle=0,width=\linewidth}}
\caption{Three-person game with imperfect competition and payoff solutions\label{payoffs}}
\end{center}
\end{figure}

\section{Conclusion}

This paper has concentrated on some of the underlying theoretical mechanics
of simulated annealing and how they relate to the trembling hand perfect
refinement of Nash equilibrium. It has been argued that the trembles that
underlie global optimization by simulated annealing are analogous to the
``mistakes'' of trembling hand perfection, in that they present a means of
moving from local equilibria. The main contribution of this paper has been
to apply simulated annealing to solve a game that is known to possess both a
perfect equilibrium and ``nonsensical'' subgame perfect equilibrium.
Preliminary results indicate a convergence to the perfect equilibrium, with
a mixing strategy occurring for two of the three players.

\end{document}